# Room-Temperature Sputtered Ultralow-loss Silicon Nitride for Hybrid Photonic Integration

Shuangyou Zhang[1], Toby Bi[1,2], Irina Harder[1], Olga Lohse[1], Florentina Gannott[1], Alexander Gumann[1], Yaojing Zhang[1], and Pascal Del'Haye[1,2,*]

[1]Max Planck Institute for the Science of Light, 91058 Erlangen, Germany

[2]Department of Physics, Friedrich-Alexander-Universität Erlangen-Nürnberg, 91058 Erlangen, Germany

*Corresponding author: *pascal.delhaye@mpl.mpg.de*

**Abstract**

Silicon-nitride-on-insulator photonic circuits have seen tremendous advances in many applications, such as on-chip frequency combs, Lidar, telecommunications, and spectroscopy. So far, the best film quality has been achieved with low pressure chemical vapor deposition (LPCVD) and high-temperature annealing (1200 °C). However, high processing temperature poses challenges to the cointegration of $Si_3N_4$ with pre-processed silicon electronic and photonic devices, lithium niobate on insulator (LNOI), and Ge-on-Si photodiodes. This limits LPCVD as a front-end-of-line process. Here, we demonstrate ultralow-loss Silicon nitride photonics based on room-temperature reactive sputtering. Propagation losses as low as 5.4 dB/m after 400 °C annealing and 3.5 dB/m after 800 °C annealing are achieved, enabling ring resonators with more than 10 million optical quality factors. To the best of our knowledge, these are the lowest propagation losses achieved with low temperature silicon nitride. This ultralow loss enables threshold powers for optical parametric oscillations to 1.1 mW and enables the generation of bright soliton frequency combs at 1.3 and 1.5 μm. Our work features a full complementary metal oxide semiconductor (CMOS) compatibility with front-end silicon electronics and photonics, and has the potential for hybrid 3D monolithic integration with III-V-on-Si integrated lasers, and LNOI.

**Introduction**

Benefiting from the maturity, low cost, and scalability of CMOS manufacturing driven by the microelectronics industry, silicon integrated photonics has enabled significant advances in telecom applications[1,2], Lidar[3], integrated photonic sensors[4] and quantum computing[5,6] over the past decade. Today, silicon photonic integrated circuits have been demonstrated with a wide diversity of optical functions via hybrid integration with different materials, such as III-V-on-Si integrated lasers, Ge-on-Si photodiodes, high-speed modulators, and (de)multiplexers. However, some intrinsic material properties of silicon pose challenges to the performance of silicon optical devices[7] in certain operation regimes. In particular, two-photon absorption of silicon at 1550 nm limits the ability of high-power handling and a high thermo-optic coefficient makes devices sensitive to temperature variations.

Silicon nitride ($Si_3N_4$) on insulator has attracted intensive interest as a dielectric material platform with ultralow propagation loss for photonic integrated circuits[8]. It has a wide transparency window covering visible to mid-IR, a relatively high nonlinearity (ten times higher than that of silica), and a low thermo-optical coefficient, as well as a CMOS-compatible manufacturing process. With its remarkable progress, it has enabled a wide range of applications in chip-scale frequency combs[9], Lidar[10], optical telecommunications[11], spectroscopy[12], on-chip delay lines[13], sensing[14], just to name a few. Among these applications, in addition to the requirement for low propagation loss, dispersion engineering and high confinement of optical mode are critical, which typically calls for film thicknesses of more than 600 nm at telecom wavelengths. There are several methods to fabricate thick $Si_3N_4$ films, such as LPCVD[15,16], plasma enhanced chemical vapor deposition (PECVD)[17–19], inductively coupled plasma chemical vapor deposition (ICP-CVD)[20], reactive sputtering[21,22] and atomic layer deposition (ALD)[23]. So far, the best optical quality of $Si_3N_4$ film is deposited by high temperature (~800 °C) LPCVD[8]. However, high film stress resulting from the LPCVD process leads to fatal cracks and prevents thick films (> 400 nm) with high optical quality[15]. As a result, several clever and sophisticated processes have been developed, such as patterning crack barriers[15], multi-step deposition[15,24,25], and a photonic Damascene process[26]. In particular, long-time high-temperature annealing (1200 °C/3 hours) is indispensable to reduce the material loss and to achieve ultralow propagation loss (~1 dB/m). This ultralow loss enables on-chip $Si_3N_4$ resonators with optical quality factors of >10 million[24,27,28]. However, the high-temperature deposition and annealing processes can cause dopant diffusion and damages prefabricated temperature-sensitive devices[21,29,30]. Hence, LPCVD $Si_3N_4$ is very challenging for applications in back-end-of-line (BEOL) fabrication processes, co-integration with front-end-of-the-line (FEOL) silicon optoelectronic circuits, and foundry compatible processes.

In recent years, low-temperature deposited $Si_3N_4$ thin films have attracted increasing interest in 3D hybrid integration of $Si_3N_4$ with other materials [18,21,29–32]. Among different low-temperature deposition methods, PECVD and ICP-CVD are widely used as low-temperature (< 400 °C), and CMOS compatible processes, providing high-thickness $Si_3N_4$ films (>1 μm) without cracks. However, the optical loss within the deposited films at telecommunication wavelengths is usually high (typically 2 dB/cm) due to the absorption of Si-H and N-H bonds[33], which poses a challenge for optical applications. To address this issue, deuterated silane ($SiD_4$) has been used as the deposition precursor, rather than conventional $SiH_4$[29,30,34], showing a 10-dB/m propagation loss. The state-of-the-art propagation loss with a PECVD process without post thermal annealing is 0.42 dB/cm, which was achieved recently by including a chemical mechanical planarization step [35]. Another potential BEOL method is low-temperature reactive sputtering of low-loss $Si_3N_4$ film with low stress [21,32,36]. Waveguide losses of 0.8 dB/cm at 1550 nm and microring resonators with an intrinsic quality factor of $6\times10^5$ have been achieved with thermal annealing at 400 °C in ambient atmosphere.

Here we report the realization of ultralow-loss $Si_3N_4$ photonic devices by reactive sputtering at room temperature. The optical propagation loss of the sputtered $Si_3N_4$ film is investigated at O-, C-, and L-band wavelength ranges and at different annealing temperatures. Directly after room-temperature sputtering, the $Si_3N_4$ film loss is 32 dB/m without thermal annealing and the intrinsic optical quality factor ($Q$) of the fabricated microresonators is 1.1 million at 1580 nm. After a CMOS-compatible annealing at a temperature of 400 °C, the propagation loss is significantly reduced to 5.4 dB/m and the resonator $Q$ is improved to 6.2 million at 1580 nm. By additional annealing at 800 °C the film quality is further improved, enabling losses of 3.5 dB/m and $Q$-factors beyond 10 million. These performance specifications are comparable to the state-of-the-art result using LPCVD films and high-temperature annealing. With the reduced optical loss, we demonstrate threshold powers for optical parametric oscillation in ring resonators down to 1.1 mW at 1310 nm. As a further verification of the optical properties of the silicon nitride films, we demonstrate the generation of dissipative Kerr soliton combs centered at both 1.3 and 1.5 microns. We believe the low-temperature sputtered $Si_3N_4$ with ultralow optical loss will have a significant impact on scalable foundry photonics and hybrid integration of 3D photonic circuits.

**Film deposition and waveguide fabrication**

Figure 1(a) shows the fabrication flow starting with a silicon wafer with 3-μm thermally oxidized silicon dioxide ($SiO_2$) layer. The $Si_3N_4$ film is sputtered at room temperature by a commercial reactive magnetron sputtering system equipped with an asymmetric bipolar pulsed DC source. P-doped silicon target with a diameter of 2 inch is used as the sputtering source. The process is using a 100-W plasma environment with a mix of Argon (Ar) and nitrogen ($N_2$). Note that the process is intrinsically hydrogen free, which avoids optical losses through N-H or Si-H bonds. The chamber pressure is controlled at 3 mTorr and the flow rate of Ar is fixed at 15 sccm. Figure 1(b) shows the refractive index of the sputtered SiNx film at 636.4 nm as a function of Ar/$N_2$ ratio, together with a dashed line indicating the refractive index of stoichiometric $Si_3N_4$ [15]. Depending on the targeted refractive index, the Ar/$N_2$ ratio can be widely tuned, resulting in the variation of the SiNx refractive index from 2.07 (silicon rich) to 2.01 (nitride rich). The refractive index is measured at 636.4 nm with a prism coupler (Metricon 2010M). For the fabricated samples presented here, the gas ratio is set to around 2.7 for stoichiometric $Si_3N_4$, and the resulting refractive index is around 2.04 at 636.4 nm with a deposition rate of 130 nm/hour. Figure 1(c) shows the top surface roughness of the sputtered $Si_3N_4$ film with a thickness of 750 nm, measured by an atomic-force microscope. The root mean square roughness is 0.34 nm, which is on the same level as that of the substrate. The sputtered $Si_3N_4$ film is used to fabricate microring resonators to investigate the waveguide propagation loss by measuring the optical quality factor of the resonances and to explore potential applications in nonlinear photonics. A negative-tone photoresist (ma-N 2405) is used to pattern optical waveguides and microrings by electron beam lithography. After developing, the patterned samples are fully etched by inductive coupled plasma reactive ion etching with an etchant mixture of $CHF_3$ and $O_2$. Figure 1(d) shows a scanning electron microscope (SEM) image of a 100-μm-radius microring resonator with a waveguide cross-section of 750 nm × 1.8 μm. Figure 1(e) shows a zoomed-in SEM image of the ring waveguide with no visible sidewall roughness. The gap between the bus waveguide and ring is varied for different coupling conditions at different wavelength ranges. After the $Si_3N_4$ etching process, piranha solution is used to remove residual photoresist. Following this step, a 3-µm thickness $SiO_2$ layer is deposited to encapsulate the samples via two steps. To fill the gap between the bus waveguides and microrings without air voids, a layer of $SiO_2$ with 400 nm thickness is deposited by ALD, followed by a 2.5-µm-thickness $SiO_2$ layer deposited by PECVD. The samples are diced for end-fire coupling to characterize their optical properties, such as optical $Q$, propagation loss, and dispersion of microresonators.

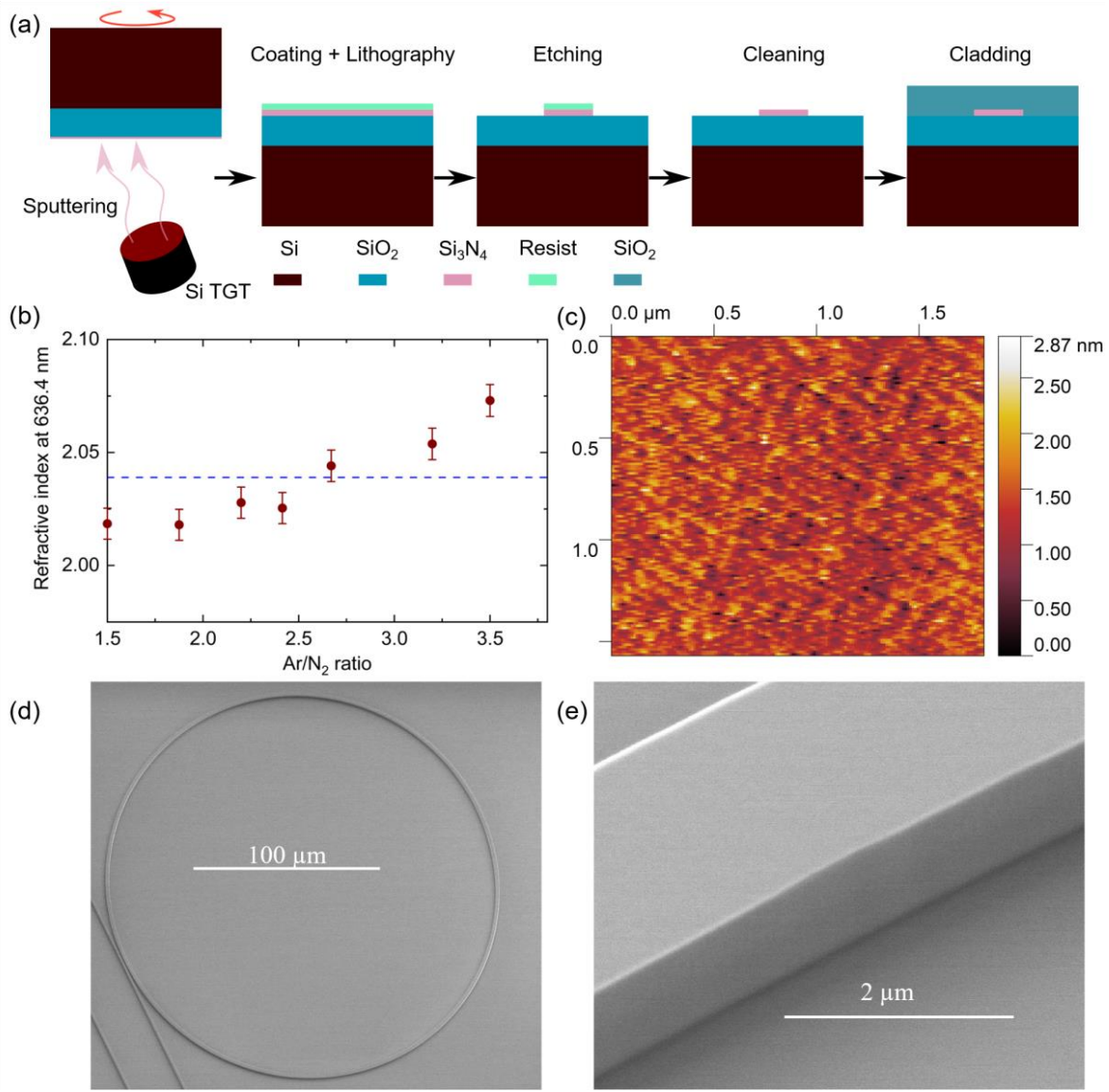

Figure 1 (a) Fabrication flow of sputtered $Si_3N_4$ photonic devices. (b) Refractive index of sputtered SiNx film as a function of $Ar/N_2$ ratio measured at 636.4 nm. (c) Roughness measurement of the top surface of sputtered $Si_3N_4$ with 750-nm thickness. (d) Scanning electron microscope (SEM) image of a 100-µm-radius microring resonator. (e) Zoomed-in SEM image of the ring waveguide.

**Characterization of the sputtered silicon nitride films and microresonators**

As shown in Fig. 2(a), two tunable continuous wave (CW) external cavity diode lasers in the O-band and the C-/L-bands are used to characterize the optical transmission of the sputtered $Si_3N_4$ microresonators and generate dissipative Kerr soliton microcombs. The polarization of the CW light is aligned by a fiber polarization controller to probe the quasi-transverse electric (TE) or quasi-transverse magnetic (TM) modes of the $Si_3N_4$ resonators. Light is coupled into and out of the resonators via two lensed fibers, with a coupling loss of less than 2 dB per facet. For characterization of the resonators, both CW lasers are set to low optical power to avoid resonance shifts induced by thermal effects and Kerr effect. The frequency of both CW lasers is widely scanned across a range of 10 THz at a speed of around 1 THz/s. To determine the laser frequency while scanning, we employ a high-precision calibrated fiber loop cavity for providing frequency markers. The transmission spectra of the $Si_3N_4$ resonators and reference fiber cavity are simultaneously recorded by an oscilloscope. Further data analysis is used to extract the optical quality factors, propagation loss, and dispersion of the resonators.

**Devices without thermal annealing**

Figure 2(b) shows a resonance spectrum of a 200-μm-radius $Si_3N_4$ resonator at a resonant wavelength of 1318.7 nm and Fig. 2(c) shows a resonance spectrum of another 200-μm-radius $Si_3N_4$ resonator at a resonant wavelength of 1578.6 nm. Both resonators are directly fabricated from a room-temperature sputtered film without post-thermal annealing. The dashed black lines in Fig. 2(b) and (c) are fitted resonance profiles using extended coupled-mode theory[37,38], showing that the resonance linewidths at 1319 nm and 1579 nm are around 300 MHz and 200 MHz, respectively, which correspond to loaded quality factors ($Q_l$) of 0.75 million at 1319 nm and 0.94 million at 1579 nm. Considering that both resonances are undercoupled, the calculated intrinsic quality factors ($Q_i$) are 1 million at 1319 nm and 1.1 million at 1579 nm, respectively. The propagation loss $\alpha$ can be estimated from the $Q_i$ using $\alpha = f_0/(Q_i \times R \times FSR)$[39], where $f_0$ is the resonant frequency, R is the radius of the ring resonator, and FSR is the free spectral range of the microresonators. Without post-thermal annealing, the room-temperature sputtered $Si_3N_4$ can already achieve a propagation loss of 0.42 dB/cm in the O-band and 0.32 dB/cm in the L-band. This propagation loss is already lower than that of annealing-free LPCVD $Si_3N_4$ film deposited at 780 °C[40]. Figure 2(d) shows the distribution of $Q_i$ of the resonances within the measurement wavelength range. Note that for both wavelength ranges, the intrinsic loss becomes larger when the wavelength gets shorter due to material loss. Similar observations are also found in LPCVD deposited $Si_3N_4$ film.

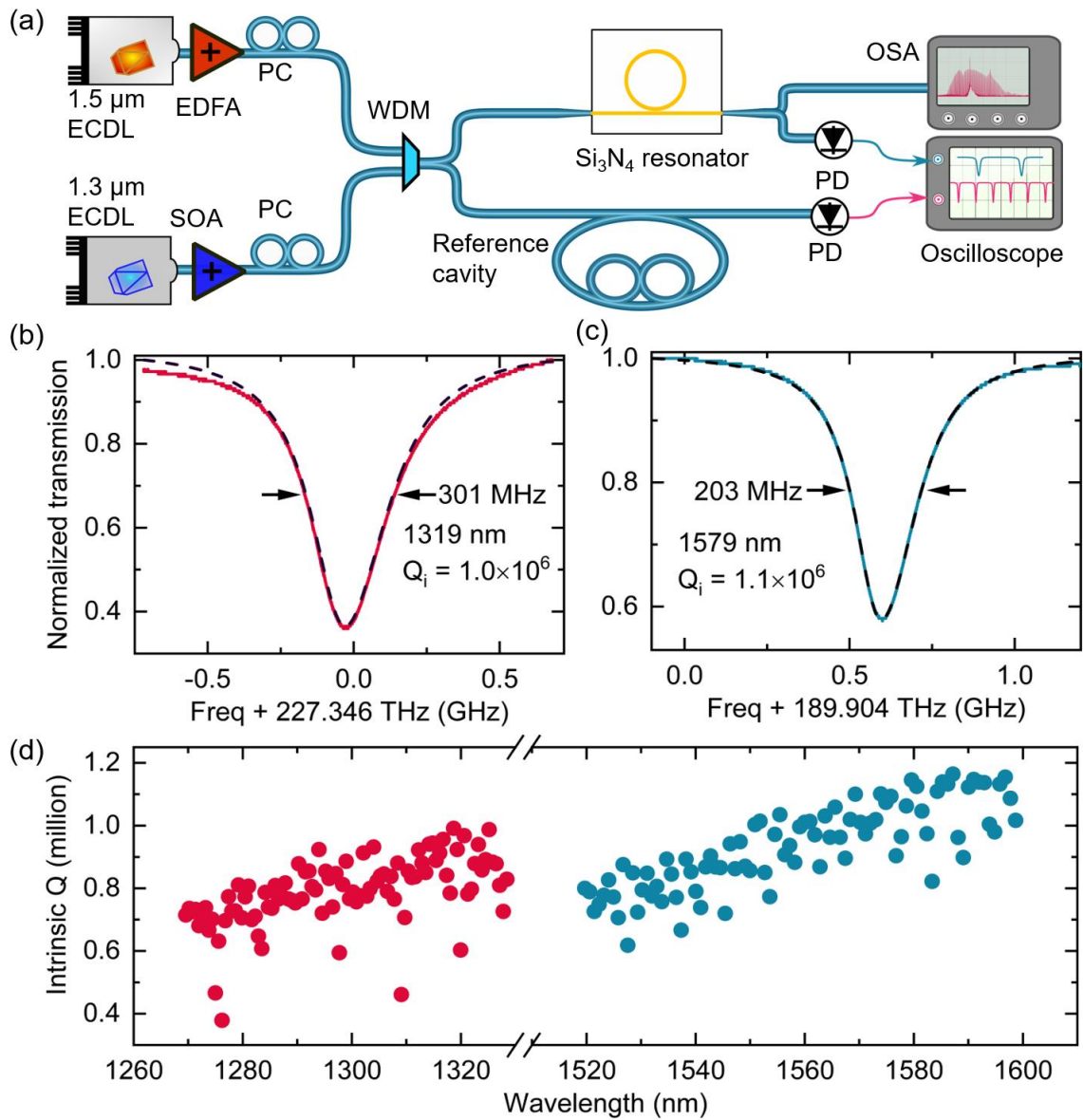

Figure 2 (a) Experimental setup for characterization of $Si_3N_4$ resonator transmission spectra, resonator linewidth, optical quality factors and dispersion. ECDL: external cavity diode laser; EDFA: erbium-doped fiber amplifier; SOA: semiconductor optical amplifier; OSA: optical spectrum analyzer; PC: polarization controller; PD: photodetector; WDM: wavelength-division multiplexer. Normalized transmission spectra of $Si_3N_4$ ring resonators without thermal annealing around 1318 nm (b) and 1578 nm (c). (d) Intrinsic *Q* measurement of all resonances within the laser wavelength ranges.

**Devices after 400 °C thermal annealing**

To reduce the material loss, the $Si_3N_4$ devices after encapsulation are post-processed with thermal annealing. First, we anneal the devices in an ambient atmosphere at 400 °C for 8 hours. 400 °C processing temperature is chosen because it is CMOS compatible and BEOL process compatible, and can be used for the integration of $Si_3N_4$ with pre-processed CMOS electronics, III-V materials, and prefabricated photonic circuitry, such as lithium niobate on insulator, Ge-on-Si photodiodes, and active silicon modulators. Figure 3(a) and (b) show the normalized resonance transmissions of two fundamental TE modes at resonant wavelengths of 1328 nm and 1610 nm of a 200-μm-radius $Si_3N_4$ resonator together with fitted line shapes (dashed lines), respectively. The loaded resonance linewidth has been dramatically narrowed to around 55 MHz after 400 °C annealing, compared to Fig. 2(b, c). Considering that the device is designed for undercoupling across the measurement range, the calculated intrinsic $Q$s at 1328 nm and 1610 nm are 4.6 million and 6.2 million, respectively, corresponding to a propagation loss of 9.0 dB/m 1328 nm and 5.4 dB/m at 1610 nm. This is approximately a 5-fold improvement compared to the propagation loss without thermal annealing. Figure 3(c) shows the distribution of the intrinsic $Q$s of the fundamental TE mode of the resonator under test covering the O-band and C+L band wavelength ranges. The intrinsic propagation loss in the O-band stays roughly constant with an average $Q_i$ of 3.8 million. We observe some low $Q$ modes, which are a result of mode crossings. In comparison, the intrinsic loss in the C-band is slightly higher than in the L-band, which we speculate is due to residual losses in the material. To the best of our knowledge, this is the lowest $Si_3N_4$ propagation loss reported with a peak processing temperature of 400 °C.

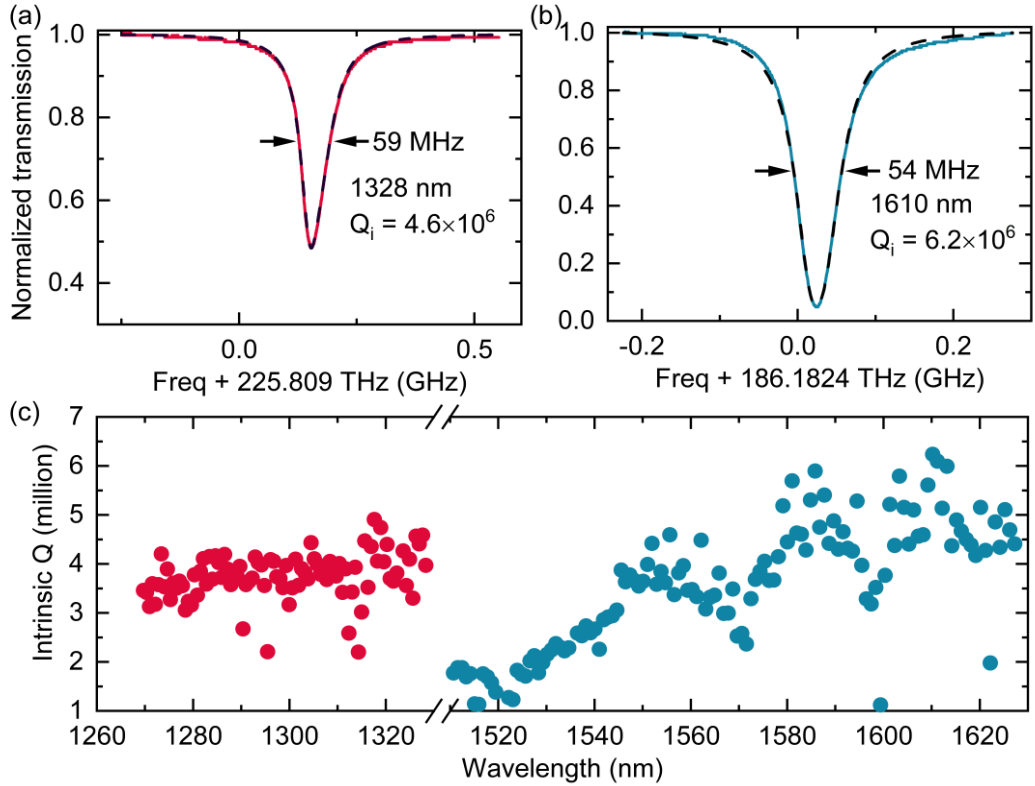

Figure 3. *Q* measurement of a 200-μm-radius Si$_3$N$_4$ ring resonator after 400 °C thermal annealing. Normalized transmission spectra of a fundamental TE mode at 1328 nm (a) and a fundamental TE mode at 1610 nm (b) of a 200-μm-radius ring resonator. (c) Intrinsic Q measurement of fundamental TE modes within the laser wavelength ranges.

**Devices after 800 °C thermal annealing**

To further reduce the material loss of the sputtered Si$_3$N$_4$ film, we anneal the devices at a higher temperature of 800 °C. This processing temperature is chosen since it is below the dopant activation temperature (1030 °C) for Si modulators[41] and the peak processing temperature (825 °C) for the fabrication of Ge-on-Si photodiodes[42]. Thus, annealing at 800 °C preserves the capability for future integration of Si$_3$N$_4$ with silicon integrated photonics. After encapsulation, the devices are post-annealed at 800 °C for 8 hours. Figure 4(a) shows a normalized resonance transmission of a 200-μm-radius Si$_3$N$_4$ resonator at a resonance wavelength of 1319 nm where the resonance is close to critically coupled but still undercoupled, together with a fitted line shape (dashed line). The loaded resonance linewidth is about 37.5 MHz. The calculated loaded and intrinsic *Q*s are 6.1 million and 9.2 million, respectively, corresponding to a propagation loss of 4.5 dB/m. The losses at 1319 nm are approximately halved compared with the losses after 400 °C thermal annealing. Figure 4(b) shows the distribution of the intrinsic *Q*s of the fundamental TE mode of the resonator in different wavelength ranges. From Fig. 4(b), we note that the intrinsic Q of some modes in the L-band is beyond 10 million, and the corresponding propagation loss is less than 3.5 dB/m. Similar to Fig. 2(c), the intrinsic propagation loss in the O-band stays nearly constant with an average intrinsic Q of 7million, except for a few low-$Q$ modes resulting from mode crossings. The intrinsic $Q$ distribution in the C- and L-bands is significantly different from the Q distribution in Fig. 3(c). After 800 °C thermal annealing, the intrinsic losses in the C- and L-bands are roughly constant as shown in Fig. 4(b). This shows that the material loss in the C-band has been effectively driven down by 800 °C annealing. The average intrinsic $Q$ crossing the C- and L-bands is 7.5 million. In Fig. 5 we compare our results with other state-of-the-art research using different low-temperature $Si_3N_4$ deposition methods. To the best of our knowledge, this is the lowest $Si_3N_4$ propagation loss reported at a peak processing temperature of 800 °C. To further verify the ultrahigh optical $Q$ of our $Si_3N_4$ ring resonator, we test the threshold optical power for Kerr-nonlinear optical parametric oscillations. The threshold power is determined by measuring the optical power of the first generated optical sideband when varying the on-chip optical power. Figure 4(c) shows experimental results when pumping an optical mode at 1312 nm of a 130-μm-radius ring resonator with a loaded $Q$ of 5.5 million and an intrinsic $Q$ of 9 million, showing a threshold power of ~1.1 mW. With an effective area of the pump mode $A_{eff}$ =0.99 μm$^2$ obtained from a COMSOL Multiphysics simulation and the Kerr nonlinear coefficient $n_2 = 2.4 \times 10^{-19}$ m$^2$/W[27], the calculated threshold power is 1.06 mW, which agrees well with the observed threshold power of 1.1 mW.

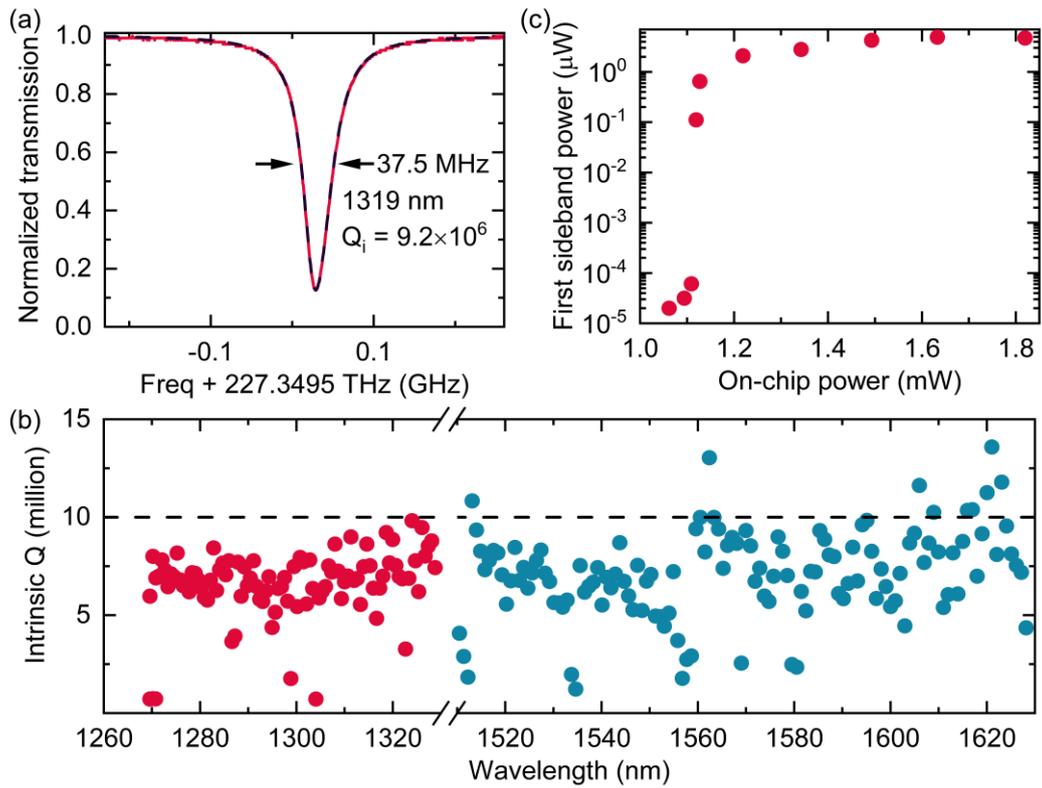

Figure 4. Q measurement of devices after 800 °C thermal annealing. (a) Normalized transmission spectrum of a fundamental TE mode at 1319 nm in a 200-μm-radius ring resonator. (b) Intrinsic Q measurement of the fundamental TE modes in the O-, C- and L-bands. (c) Optical parametric oscillation threshold power measurement for a 130-μm-radius ring resonator.

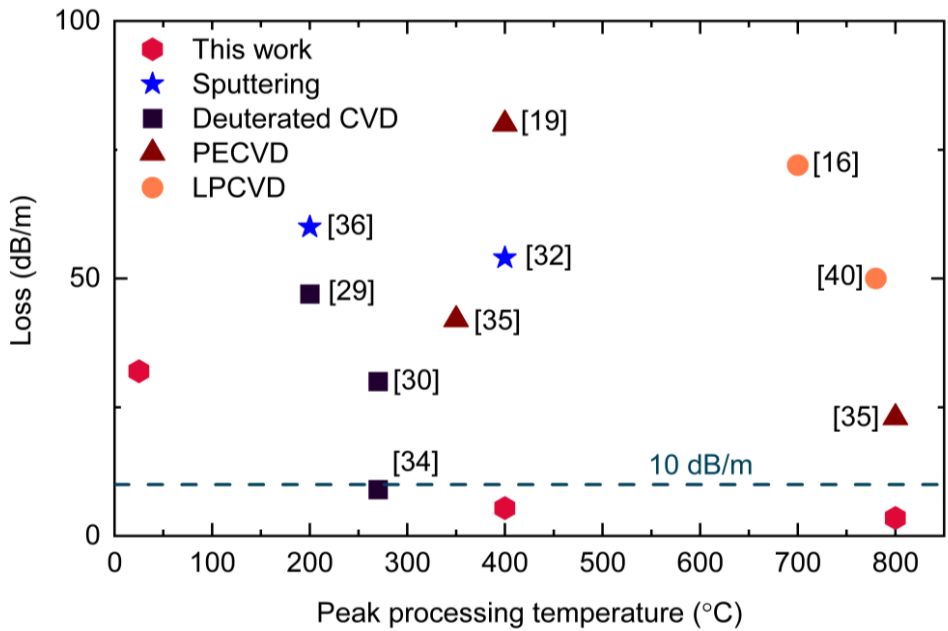

Figure 5 Propagation loss comparison with other state-of-the-art research based on low-temperature deposition techniques, annealing-free LPCVD, PECVD, sputtering, and deuterated silicon nitride based on ICP-CVD.

**Soliton frequency comb generation in sputtered Si$_3$N$_4$**

In this section, we use the fabricated Si$_3$N$_4$ microresonators for soliton frequency comb generation. We first measure the dispersion of a 100-μm-radius resonator with a cross-section of 750 nm × 1.8 μm. The resonance frequencies of a mode family in a resonator can be described with a Taylor series as[26,43]

$$\omega_\mu = \omega_0 + D_1\mu + \frac{D_2}{2!}\mu^2 + \ldots, \\ = \omega_0 + D_1\mu + D_{\text{int}}(\mu) \tag{1}$$

where $\mu$ is the mode number offset from the center mode at $\mu = 0$ and $\omega_\mu$ are the resonance frequencies. $D_1/2\pi$ is the FSR of the resonator at the center mode ($\mu = 0$), and $D_2$ is the coefficient of second-order dispersion. $D_{\text{int}}$ is the integrated dispersion, depicting the deviation of resonance frequencies from the equidistant grid spaced by $D_1$. Figure 6(a) shows the measured integrated dispersion profile (blue circles) at a center mode of 1310 nm together with a second-order polynomial fitted curve (blue curve). The dispersion profile shows anomalous dispersion with mode crossings at 1280 and 1326 nm and is suitable for bright soliton generation. We test this Si$_3$N$_4$ resonator for the generation of a bright soliton frequency comb by pumping a mode at 1310 nm. The experimental setup is shown in Fig. 2(a). For characterizing the resonators, both lasers operate at very low optical power. During soliton generation, we increase the power of both lasers to ~100 mW. To overcome the thermal effect of the microresonator, the 1.55-μm laser is simultaneously coupled into a resonance to thermally stabilize the intracavity power [44]. Figure 6(b) shows the transmission traces when scanning the 1.3-μm pump laser across the resonance from the blue- to red-detuned side. Different soliton steps, corresponding to different soliton numbers, are observed in the red-detuned regime (green shaded area)[44,45]. Using the passive temperature stabilization with the auxiliary laser at 1.55 μm, bright solitons at 1.31 μm can be accessed by manually tuning the laser frequency into the red-detuned soliton regime. Pumping with 80 mW on-chip power at 1.3 μm wavelength leads to the generation of a bright soliton, shown in Fig. 6(c) together with a fitted sech$^2$ envelope (black dashed). Note that the mode crossing at 1326 nm (c.f. dispersion profile in Fig. 6(a)) induces a dispersive wave (marked with an arrow) at 1326 nm in Fig. 6(c). In addition, we also demonstrate a bright soliton frequency comb at 1.55 μm. Figure 6(d) shows an optical spectrum of a bright single soliton when pumping a mode at 1574 nm. In this case, the 1.3-μm laser is used as the auxiliary laser to thermally stabilize the resonator.

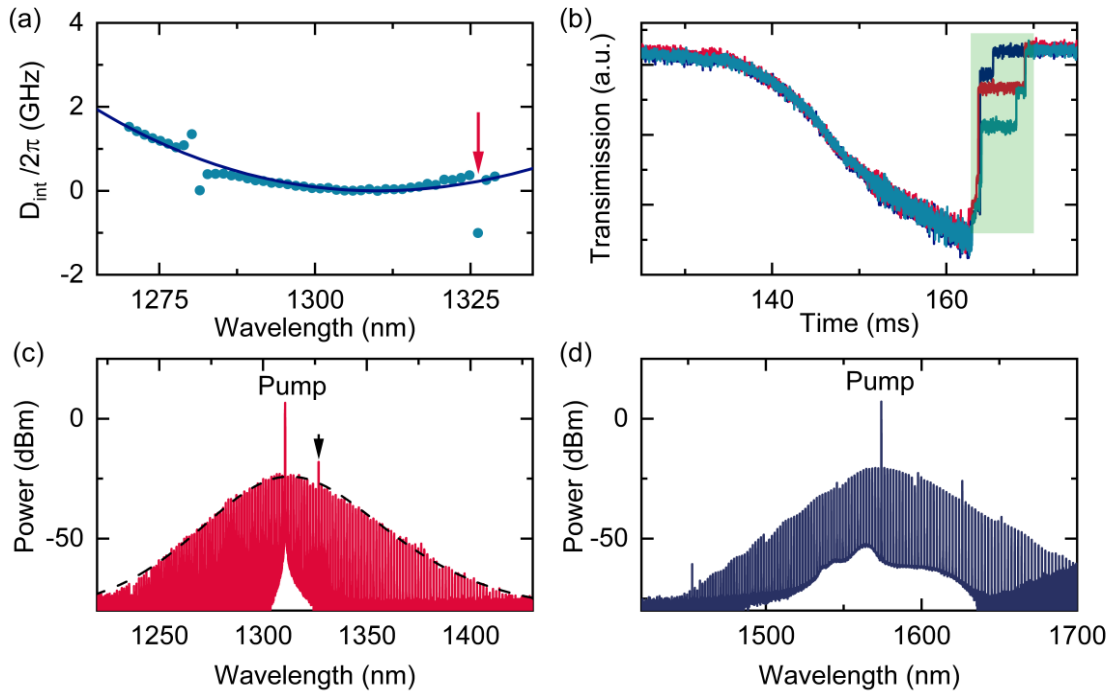

Figure 6 (a) Measured integrated dispersion profile (blue circles) at a pump wavelength of 1310 nm together with a second-order polynomial fit (blue trace). (b) Measured transmission traces when scanning the 1.3-μm pump laser frequency from the blue-detuned to the red-detuned side of the resonance. (c) Optical spectrum of a soliton frequency comb generated in the sputtered $Si_3N_4$ resonator pumped at 1310 nm and a $sech^2$ envelope fit (black dashed trace). (b) Optical spectrum of a soliton frequency comb pumped at 1575 nm.

**Conclusion**

In summary, we report the fabrication of ultralow-loss $Si_3N_4$ photonics using room-temperature reactive sputtering. The propagation loss is measured at 32 dB/m without post-thermal annealing. After thermal annealing at a CMOS-compatible processing temperature (400 °C), the film loss can be dramatically reduced to 5.4 dB/m, enabling the fabrication of $Si_3N_4$ ring resonators with an intrinsic $Q$ of 6.2 million in the L-band. This process for ultralow loss integrated $Si_3N_4$ devices enables fully monolithic BEOL integration with pre-processed CMOS electronics, silicon integrated photonics[7], and LNOI[46]. After further thermal annealing at 800 °C, the film loss is further reduced to 3.5 dB/m corresponding to attainable optical quality factors of microresonators beyond 10 million. This performance is also the lowest $Si_3N_4$ propagation loss reported at a peak processing temperature of 800 °C. We believe that our low-temperature sputtered $Si_3N_4$ is a significant step toward the hybrid monolithic integration of silicon nitride with silicon photonics.

## Data availability

The data that support the findings of this study are available from the corresponding authors upon reasonable request.

## Acknowledgements

This work is supported by European Union's H2020 ERC Starting Grant "CounterLight" 756966; H2020 Marie Sklodowska-Curie COFUND "Multiply" 713694; Marie Curie Innovative Training Network "Microcombs" 812818; and the Max Planck Society. The authors would like to thank Prof. Victor Torres-Company and Krishna Twayana for helpful discussions.

## Author contributions

S.Z. and P.D. conceived the project. S.Z. with input from T.B. and Y. Z. designed and fabricated the microresonators. S. Z., I. H., and A. G. did the Si3N4 sputtering. O. L. did the electron beam lithography. F. G., A. G., and S. Z. completed the SiO2 cladding. S.Z. performed the measurements. S.Z., T.B., and P.D. analyzed the data and wrote the manuscript with input from all other co-authors.

## Additional information

Correspondence and requests for materials should be addressed to P.D.

## Conflict of interest

The authors declare that they have no conflict of interest.